# EdgeSphere: A Three-Tier Architecture for Cognitive Edge Computing

Christian Makaya, Keith Grueneberg, Bongjun Ko, David Wood, Nirmit Desai, Xiping Wang
IBM Research
Yorktown Heights, NY, USA

*Abstract*—Computing at the edge is increasingly important as Internet of Things (IoT) devices at the edge generate massive amounts of data and pose challenges in transporting all that data to the Cloud where they can be analyzed. On the other hand, harnessing the edge data is essential for offering cognitive applications, if the challenges, such as device capabilities, connectivity, and heterogeneity can be overcome. This paper proposes a novel three-tier architecture, called *EdgeSphere*, which harnesses resources of the edge devices, to analyze the data *in situ* at the edge. In contrast to the state-of-the-art cloud and mobile applications, EdgeSphere applications span across cloud, edge gateways, and edge devices. At its core, EdgeSphere builds on Apache Mesos to optimize resources usage and scheduling. EdgeSphere has been applied to practical scenarios and this paper describes the engineering challenges faced as well as innovative solutions.

## I. Introduction

By 2020, it is predicted that there will be as many as 25 Billion edge devices. Applications of leveraging such devices span across industries: manufacturing, connected home, smart cities, industrial process, and healthcare to cite just few. The Internet of Things (IoT) is defined as the collection of everyday machines and smart objects that are now embedded with sensors and/or actuators and that can communicate over the Internet. One of the biggest challenges posed by such adoption is the exponential growth of data generated by these devices. It is estimated that the data growth is outpacing network bandwidth improvements by a factor of two. While insights extracted from such data are extremely valuable to businesses and users alike, the trends above indicate that having to transport such high volumes of data over the network will be increasingly challenging. In fact, it is estimated that more than 80% of the data remains untapped – not analyzed, stored, or transported, due to these constraints.

The state-of-the-art paradigms for analyzing data generated at the edge are two-tiered: a centralized tier of cloud services and a distributed collection of edge gateways, mobile and IoT devices connected to the cloud. In such paradigms, most of the computation and analytics occur at the cloud tier, requiring transport of a huge amount of data. The bandwidth limitations, higher latency, and the cost of transporting the data to the cloud are well understood challenges in the edge and fog computing literature. The edge computing approaches such as cloudlets propose to localize as much as possible computation, applications, data storage, and other computing services on the edge tier. Although these approaches are a step in the right direction, they suffer from the limitations of the two-tier paradigm described next.

In the two-tier architecture, most of the IoT devices are deployed in private networks and are owned by independent entities, e.g., homes, hospitals, and industrial plants. Two-tier approaches require a one-time registration assuming a complete trust between these devices and a central cloud service or even an edge gateway, introducing a major security and privacy risk. Furthermore, many of the IoT devices have significant computational, sensing, and communication resources, e.g., smart phones, tablets, autonomous vehicles, and drones. Two-tier approaches miss the opportunity to leverage such resources by focusing on cloud and edge gateways exclusively for all computation. Moreover, the IoT environments are highly dynamic due to the mobility of the devices as well as the changes in the available resources, e.g., low battery. The state-of-the-art is missing resources and context-aware mechanisms for the deployment of analytics such that cloud, edge gateways, and edge devices resources can be allocated optimally. Lastly, in the current deployment approaches, the application topology as well as the binding to physical target devices is fixed at the time of the deployment. Given the dynamic nature of IoT environments, this limits the class of applications that can be supported. The drawbacks of the two-tier architecture have motivated the proposal of the three-tier architecture as illustrated in Fig. 1.

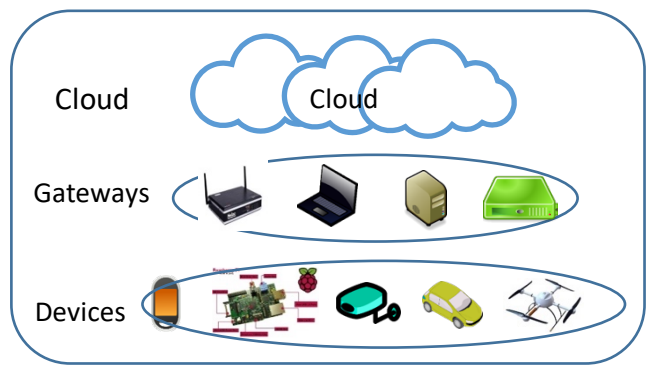

Fig. 1. Three-tier architeture for Internet of things (IoT).

The initial proposal of such architecture was made in ETSI M2M specification [1]. This architecture is composed of M2M

Device and Gateway Domain (where edge devices and edge gateway are deployed), and M2M network domain (represented by the core network and cloud data center). The IoT applications domain can be considered as part of the cloud in the ETSI M2M architecture. With the three-tier architecture, analytics services can be deployed at the edge gateway where raw data processing (aggregations, filtering) can be done and possibly generate actionable events. The edge gateway adds control and improves management closer to the edge rather than sending the raw data to the cloud (which leads to a higher bandwidth consumption and latency). Only the insights and possibly filtered data are sent to the cloud for reporting and further analysis.

This paper proposes a novel three-tier architecture, called EdgeSphere, to address the limitations identified above. EdgeSphere automates the orchestration and deployment of distributed IoT applications while guaranteeing efficient usage of resources, scalability, and fault-tolerance. Specifically, EdgeSphere enables context-aware and resource-aware deployment of IoT applications across the three tiers namely cloud, edge gateways, and edge devices. Edge devices, gateways, and cloud participate in EdgeSphere by installing an agent matching their logical role and platform. Edge devices such as smartphones may adopt the roles of edge devices as well as edge gateways, even at the same time. The agents encapsulate the protocols necessary for the communication between edge gateways and cloud as well as edge gateways and edge devices. Edge gateway agent automatically discovers edge device agents and the edge device agents report available resources to the edge gateways, which in turn aggregate such reports and communicate to the cloud.

No pre-existing trust relationship needs to exist between edge devices and edge gateways, or between edge gateways and the cloud tier. Edge devices communication may switch from one edge gateway to another. Similarly, edge gateways may communicate with multiple cloud services. Edge device owners can control what resources and data are shared with edge gateways through policy-based management. Another key contribution of EdgeSphere is the resilient management of tasks failures. Ones of the key goal of EdgeSphere is to enable deployment of advanced analytics applications at the edge. These analytics applications may require a variety of runtime environments, e.g., Java, Python, JavaScript, Docker, Machine Learning libraries, rule engines, Raspberry Pi, etc. Hence, EdgeSphere deployment is runtime-agnostic. The IoT applications describe their dependencies in a manifest and those are deployed along with the application itself. EdgeSphere is built on and adapts Apache Mesos [2, 3], which is a platform for fine-grained sharing and management of resources (e.g., memory, CPU, ports, disk, GPU) in data centers.

As a motivating example and scenario, a digital marketer may have a cloud service that determines what offers target what class of users. Retail storeowners may run edge gateways on their premises, communicating with multiple digital marketers. The marketers may deploy analytics representing specific queries to identify target consumers, e.g., "customer looking for apparel", or "customer who visited three other stores in the last hour". Consumers with smartphones playing edge device roles may visit one or more retails stores, sharing available resources. Upon discovering the consumers and their device resources, edge gateways deploy the analytics to the edge devices and only the result of the analytics are sent back to the gateway. Based on such results, the edge gateway and the cloud may decide to show one or more offers to the consumer. EdgeSphere, is a first step towards enabling such scenarios.

The remaining of this paper is organized as follows: Section II describes the related work; Section III discusses issues in Apache Mesos when used for edge computing and support of IoT applications. Section IV presents the proposed EdgeSphere framework. The prototype and use cases are described in Section IV while Section V concludes the paper.

## II. RELATED WORK

Several technologies have been developed for an efficient management of edge devices. The state-of-the art around fog and edge computing provides a framework for deploying IoT applications using two-tier based architecture. Sapphire [4] is a distributed programming system that simplifies the development of mobile and cloud applications. It handles applications' specific distribution issues such as replication, caching, and load-sharing. Sapphire gives a fine-grained control to application developers over scalability, availability, and performance. The proposed architecture is based on two-tiers, i.e., the mobile devices are directly connected to the cloud. Moreover, Sapphire does not address data management needs. CloneCloud [5] is a code-offloading system that automatically partition applications and execution runtime and offload intensive computational tasks from mobile devices onto a device's clone running in the cloud. By offloading application's tasks to the cloud, CloneCloud cannot guarantee the context accuracy for IoT applications.

Apache Edgent [6] is a programming model and runtime for edge devices which enables analysis of streaming data on the edge devices. By using Edgent, the amount of data transmitted to a back-end analytics server in the cloud can be reduced as well as the stored data. EdgeSphere has been used to deploy Edgent's applications on edge devices (e.g., Android phones and tablets) and deliver analytics on various sensors data such as accelerometer, barometer, gravity, gyroscope, and motion. Most of these prior works only solve a subset of deployment issues and challenges faced by IoT applications across distributed environments. Moreover, they have been designed for high-end infrastructures and devices.

Apache Mesos [2, 3] is a platform for resources management in data centers. It allows fine-grained sharing of resources (e.g., memory, CPU, disk, GPU) across frameworks and allows frameworks to achieve data locality by taking turns reading data stored on each machine of a cluster. A Mesos master decides how many resources to offer to each framework based on defined policy by the infrastructure manager or an organization, while the framework's scheduler decides which resources to accept and which computations to run on them. Fig. 2 shows the main components of the Mesos platform:

- *Master*: it manages workers, offers resources to frameworks, tracks resources available on the agents. The offer is a list of available resources on multiple agents;

- *Agents*: are worker nodes running on each cluster node and advertise resources to the master and run applications or tasks.
- *Framework*: defined by two entities, *Scheduler* and *Executor*. The scheduler receives deployment requests and assigns execution to the agents based on resources requirements and availability. It registers to the master to receive resources offers. While the executor is the process launched on the worker nodes to run and execute the framework's tasks.

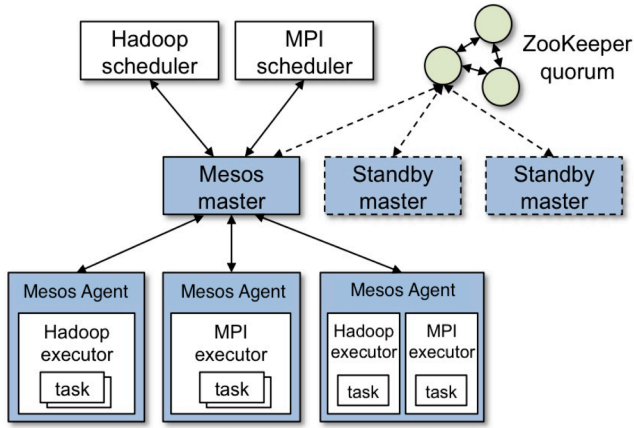

Fig. 2. Apache Mesos Architecture Diagram [3].

Mesos has attracted lot of interests and there are multiple frameworks available either as open source or commercials. For more details about Mesos project, the reader can refer to [3]. However, Mesos cannot be used out-of-the box for IoT applications and architecture due to several issues and challenges described in Section III. But, by using concepts from Mesos, we can take advantage of efficient resources management, isolation, scalability, and fault tolerance while supporting various IoT applications deployment requests across cluster of edge devices, edge gateways, and cloud. Mesos provides a stronger foundation to EdgeSphere. Below, we describe two main Mesos frameworks widely adopted.

Apache Marathon [7] is a meta-framework that can be used to manage and orchestrate applications as well as other Mesos frameworks. It automatically handles failures (hardware or software) and ensures that an application is "always on", i.e., if an application fails, Marathon will restart it on another Mesos agent. Apache Aurora [8] is another services and applications scheduler that runs on top of Mesos for long-running services that takes advantage of Mesos capabilities such as scalability, fault-tolerance, and resources isolation. When machines experience failure, Aurora can reschedule the jobs onto healthy machines. Fig. 3 shows the different components of an Aurora cluster. *Aurora scheduler*: the scheduler is the primary interface for the work running in the cluster. It manages jobs in Mesos and interfaces with the master to control the cluster. *Aurora executor:* the executor (a.k.a. Thermos executor) is responsible for carrying out the workloads described in the Aurora domain-specific language. The executor is what actually executes user processes and it performs health checking of tasks. Moreover, it registers tasks in ZooKeeper for the purposes of dynamic service discovery. *Apache ZooKeeper* is a distributed consensus system, used for reliable election of the leading Aurora scheduler and Mesos master. *Aurora observer:* the observer provides browser-based access to the status of individual tasks executing on the worker machines.

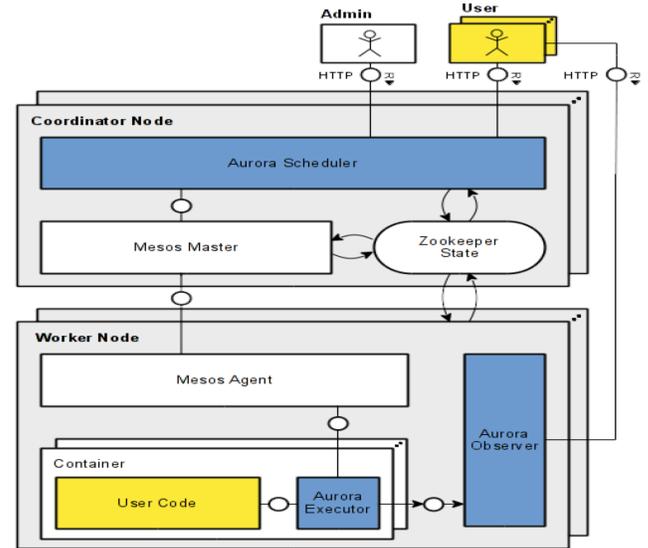

Fig. 3. Components of Aurora cluster [8].

III. CHALLENGES FOR EDGE APPLICATIONS MANAGEMENT

*A. Data Locality*

Many IoT scenarios require the processing of data to be performed close to the edge devices (either on the device or on the gateway). Mesos scheduler is not aware of and does not provide a primitive for data locality, only the host resources (e.g., CPU, memory, disk, GPU, application port number). This may cause processing to occur far from the device which would translate in high energy expenditure and transmission cost of the data across the network. Mesos provides the capability of schedulers rejecting offers from the master, and a delayed scheduling approach would have the scheduler wait until an offer that satisfies data locality is submitted. This approach introduces a delay and efficiency trade-off, i.e., to achieve data locality some requests may take too long to start.

Mesos is built for horizontal topology in data centers, where all nodes are located in the same network and administrative domain, and it is focused on the scheduling of physical hosts resources (e.g., CPU, memory, disk, and GPU). The deployment of Mesos for the support of applications into edge devices may requires proxy or virtual agents. In IoT applications, it is important for resource models and allocation to consider the edge devices attributes such as sensors, connectivity, logical capabilities, and geo-location. In order to consider cognitive and machine learning for IoT applications there is a need for a new framework and scheduling approaches that considers various resources and edge attributes such as data locality, geo-location of components, temporal constraints, privacy, and connectivity.

*B. Failure Handling*

Failures in IoT environment can occur very often due to nodes disconnections (either the agent's process is aborted or the device turns off) or due to link disconnections (due to wireless channel variations and device mobility). Link disconnections can be transient. If the agent detects a failure, it will notify the corresponding framework scheduler to do task failure handling. Typically, the framework will restart a task on a new agent node, if it accepts the resource offer from the master. There is one exception in which the compute node is alive, but the agent process dies. Mesos implements an agent recovery feature that allows executors/tasks to continue running even when the agent process exits or fails. When a task is being executed, the agent checkpoints metadata about the task to a local disk. If the agent process fails, the task continues running and when the master restarts the agent process because it is not responding to messages, the restarted agent process will use the check pointed data to recover the state and to reconnect with executors/tasks. This feature is very useful in IoT due to potential failures or link disconnections.

*C. Transient or Permanent Disconnection*

Mesos does not distinguish between transient or permanent disconnections. What happens in case of a temporary link disconnection? If Mesos master does not see the status update from an agent it assumes that the node has failed, the agent has failed and hence the task has failed. In this case, it will coordinate with the framework and try to re-deploy a new agent running the task. Meanwhile the agent continues execution in the data plane doing processing or sending data to a destination per task specifications. This issue may be dealt at the master side using detection of transient or permanent failure through some monitoring technology. One idea for detection of transient vs. permanent would be to track link disconnections on master's side by monitoring the timing of agent failures. If many agents fail simultaneously this could be an indication that the master has temporarily moved to another location. Another idea to detect disconnection on agent side would be for the master to keep history of locations where failure happened as well as the time it took for the recovery. Once the detection is achieved, there can be two ways of acting. First, using appropriate timeout mechanisms. Apart from adaptive timeouts, another type of measure would be to use task replication upon re-deployment. A related question is what happens to the agent when it disconnects from the master (its reported messages are lost). Most likely it will continue execution and will not abort. The agent should be built in soft state manner. If the master supports transient disconnections the agent may infer what the master will do. If it infers that the master will wait, otherwise it can kill its own task because the master will think that it is dead.

*D. Online Performance based on Metric Allocations*

When there are multiple frameworks, Mesos allocator will be involved. When there is a need to schedule multiple tasks, the scheduler of the framework is used. One issue is at what level should the isolation take place? Mesos only provides isolation of CPU/memory resources among frameworks using Linux containers. One natural point is the resources among virtual agents at the gateway. It makes sense to manage resources at the gateway. The aggregation and compute of intensive operations are done at the gateway and that is where we need isolation of the resources and where it can be done. In addition, since there exists a virtual agent for each device in the gateway, the resources consumed by this agent should be considered in addition to the device constraints. This can be achieved by reporting the agent's key performance indicators (KPI) in addition to the device's KPI to the master. Another scenario that needs resources management at the gateway is when tasks are ran standalone in virtual agents in the gateway.

However, for edge devices there exist no isolation mechanisms. One issue is how to cope with the fact that there are no isolation mechanisms for tasks running on edge devices. In this case, only "elastic" service can be offered, i.e., an application can run at the expense of existing applications. One solution is to use a metric other than CPU/memory that is communicated to the master. This metric will indicate how much a new task will suffer given the current state of the edge device. Thus, instead of isolating tasks, a metric which is a function of what tasks are there is computed and this metric is communicated back to the framework scheduler or master's allocator. The metric will show whether the application can be executed in this device or not. The metric can also be used to provide preference of one edge device over another.

Another solution would be to use model-based estimation and agent reports this model when it starts. Then the master passes that information to the framework scheduler and keeps track of the resource. The problem is that Mesos agent does not report the resource availability during operation. It reports whether the task is running or not. At the beginning, when the agent starts, it reports to the master the physical resources. Although each framework has monitoring capabilities where nodes report their KPI, it would need to bypass Mesos master and agent, leading to missed probe on each machine. One fundamental idea here is not to have the agents report their KPI but have the master probe them when there is a need for a task allocation. However, not all agents may be able to respond to such probe, because they do not know about the task yet. Thus, the need of a *scout probing task* approach which is used to probe one or more agents before starting the actual task. Hence, the master will send the offer to the framework. This intermediate step may be complicated because there may be multiple competing frameworks. However, it is feasible if the frameworks are coordinated to take the provided offers based on probing and Mesos master is not going to give it to anyone else unless it is released.

## IV. EGDESPHERE FRAMEWORK

EdgeSphere is designed to allow the abstraction of resources across the three layers (edge devices, gateways, and cloud) as a collection of resources pool. In other words, it provides a generic set of capabilities for edge (IoT) application deployment. Furthermore, it allocates the resources pool to the distributed and composite applications. The resources are of different types: compute (e.g., CPU, memory, I/O, disk, GPU), network (e.g., bandwidth), storage, location and characteristics of data, and platform requirements. Moreover, with EdgeSphere, IoT applications can be developed independently of the run-time and underlying network technology. At the deployment, EdgeSphere validates the operational requirements of the applications such as data locality, geo-location of devices, runtime dependencies, and network connectivity. EdgeSphere is agnostic to the IoT devices since it doesn't require changes in the devices. With optimized resources management from Apache Mesos, EdgeSphere allows IoT applications to scale automatically based on the application constraints and business logic. The key differentiating capabilities of EdgeSphere are:

- Federated edge resource management: agent managing edge resources acting as the proxy to higher-level agents, i.e., hierarchical or in peer-to-peer mode for resources sharing.
- Efficient resource management, isolation, scalability, and fault tolerance while supporting various IoT applications deployment requests across cluster of edge devices, edge gateways, and cloud.
- Dynamic context-aware resource discovery and applications deployment.
- Cognitive scheduling: learn optimized resources allocation and applications topology and dependencies via monitoring and persistent state management.

EdgeSphere architecture is illustrated in Fig. 4. The edge gateway can be dynamically reassigned in case of failures (as supported by the agent recovery in Mesos). EdgeSphere supports secure communications across edge devices, gateways, and cloud. In fact, a VPN tunnel can be setup between components of different layers. In EdgeSphere, the edge gateway is a logical role, typically assigned to one of the edge devices with sufficient power, computing resources, connectivity to cloud, and limited mobility relative to the location. Moreover, the master collects the resource models (e.g., CPU, memory, attributes, etc.) from the agents (running on edge gateways) and builds the abstract model of resource pool. The IoT applications (i.e., Mesos framework) make a request for resources based on the logical resource models. Upon the receipt of the request, the master offers the resources pool matching the requests in the response. If the application accepts the resource offer, then the master allocates the resources to the application tasks. This flow means that EdgeSphere matches the IoT applications dependencies with edge resources and finds the optimal deployment across edge devices, edge gateways, and cloud. Furthermore, EdgeSphere can dynamically move components (e.g., analytics services) across devices, gateways, and cloud.

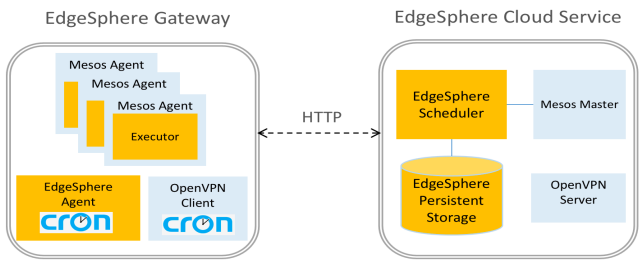

Fig. 4. EdgeSphere Architecture.

### A. Contextualization of Resources Management

In many IoT applications, it is highly desirable to deploy the application's tasks at the edge devices which are within the particular *context* required by the application, in order to efficiently utilize the limited resources at the edge. For example, suppose an application requires that the data from a set of sensors in a certain geographic region shall be processed by an analytics task at the edge devices at a certain time of the day. In such a case, the "context" in which the analytics tasks are to be deployed is the specified spatio-temporal boundary, and it is imperative such constraints be met when it comes to managing the resources for task deployment at the edge.

While such spatio-temporal conditions are some of the more widely-used contexts that can be used in the edge resource management, the general definition of the contexts could be arbitrary. For instance, one can employ an event-driven notion of the context, in which the detection of certain event (detected, e.g., by an analytics task already deployed at the edge) could determine the beginning of the particular context that the application tasks shall be deployed in. Another commonly used context is the availability of particular types of sensing devices to perform certain tasks at the edge device.

EdgeSphere provides the interfaces and mechanisms for such contextualization of the resource management for application tasks deployment. More specifically,

- EdgeSphere's edge agents constantly discover the available resources and the environmental parameters of the available edge devices, and report these to the EdgeSphere resources scheduler as a set of attributes.
- When the applications submit to the EdgeSphere's scheduler the tasks to be deployed at the edge, they specify the desired or required conditions for the task deployment, in terms not only of the amount of the resources but also of the attributes of the resources.
- EdgeSphere's scheduler performs the matching of the application's requirements to the edge resources. While EdgeSphere has a built-in, default scheduler that performs such a matching, based on the resources availability and resources attributes in a set of key-value pairs, it also provides a "hook" to allow a custom method of how such matching shall be performed can be instead utilized.

### B. Dynamic Edge Resources Discovery

In contrast to data center environments, the resources at the edge are typically heterogeneous (in their types and capabilities) and characterized by high churn rate. From the resources management point of view, it is a challenge to keep track of such dynamic and heterogeneous resources at scale. EdgeSphere addresses this challenge using a hierarchical resources management approach, in which the availability of the resources at lower tier (e.g., the device layer) is exposed to the higher tier (e.g., the scheduler in the cloud) only in an aggregated form, where the aggregation takes place in the middle tier (e.g., the gateway). This way, the resource scheduler in the cloud performs its job only based on the aggregated view of the resources under each edge gateway, with the details of the dynamism and heterogeneity of the individual edge resources hidden from the scheduler. The edge gateway instead performs its own resource management function based on the specific resources under its control, yet in a much smaller scale than what would need to be done when a single, central scheduler in the cloud takes care of all edge resources. The gateway can also exercise a certain level of autonomy in terms of how to aggregate the edge resources; for example, it may decide that only a portion, say, 50%, of the currently available edge resources be exposed to the central scheduler, so that it can spare the remaining resources in case of the edge device failures and cope with them on its own.

EdgeSphere Agent is based on two components: Gateway Discovery Agent and Agent Launcher. The *Gateway Discovery Agent* is used to discover and register edge devices. The discovery allows the edge devices to learn the capabilities of its peer devices. For example, if a task requires a camera and a microphone, but these two sensors are available on two different devices, the discovery procedure helps to know which device can provide the needed resources and capabilities for the task. The registration request carries information about the resources on the edge devices and capabilities (e.g., OS, CPU, memory, sensors, location, network interfaces, etc.). Once an edge device is registered, the *Agent Launcher* starts a Mesos agent on the edge gateway. When the Mesos agent has been started, it registers to the Mesos master and starts reporting the logical resources model to the master. These abstract and logical resources correspond to the resources and capabilities of the edge devices. Fig. 5 shows the call flow for deploying an IoT applications with EdgeSphere. In EdgeSphere, Mesos agent acts as a proxy agent for the edge devices which are the physical devices where the tasks will be executed. By proxying the edge devices, we can provide Mesos support for IoT applications where edge devices may not have the capabilities to fully support Mesos (e.g., constrained IoT devices). The services and resources discovery are based on our previous work called distributed state machine (DSM) [9] for the peer discovery.

DSM is a framework for sharing state across distributed applications and allowing collaborative monitoring of nearby devices to secure data flows (i.e., negatively observed nodes are avoided). In DSM, applications collaborate to solve distributed problems (e.g., network management, sensors data collection and processing, distributed state, etc.). It supports policy enforcement over shared data, i.e., filtering and access control on inbound and outbound data and queries. Generalized capabilities to allow pub/sub type data access across DSM nodes is key feature of DSM and is used to enable resource discovery in EdgeSphere.

The discovery agent can be disabled in EdgeSphere for the scenarios where Mesos agents can directly run on edge devices and no proxy agent is required between the edge devices and the IoT devices (e.g., sensors). In this case, the agents are configured to register to the master when they start. When a new edge device is up, the DSM agent collects local, shared, and previous existing observations into a list. The new opinions are computed as the average of all observations of a node. After this computation, the local observations are then sent to the neighbor nodes. The discovery agent periodically inspects the state of the nodes in the topology and can query the shared data via a remote peer application or continuously query via remote peer subscription.

With EdgeSphere, the integration of Mesos has been done by implementing a *Mesos Scheduler* for service management and a *Mesos Executor* for tasks management and execution. Both are defined as abstract classes that can be extended for various Mesos frameworks. This simplifies the design and development of new edge applications or Mesos frameworks. A dashboard is also provided as the user interface for the deployment of the IoT edge applications and to view the cluster nodes, the status of tasks, logs, etc. The master node can be deployed in the cloud (e.g., IBM Bluemix) or on-site close to the edge network.

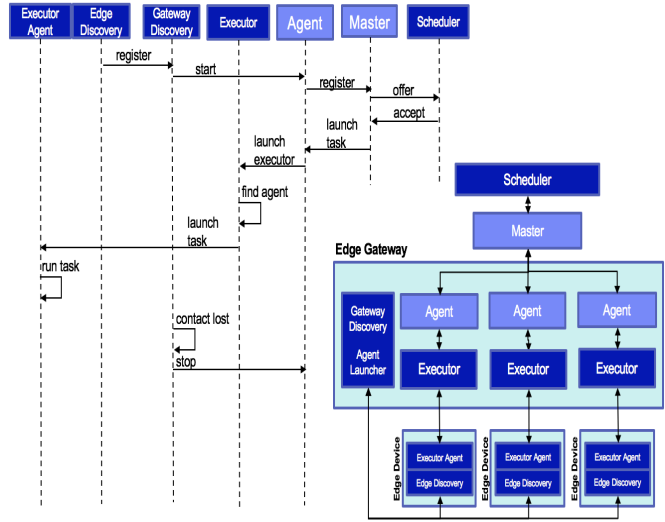

Fig. 5. Messages Flow in EdgeSphere.

*C. Scheduling and Executing Tasks*

EdgeSphere scheduling service is a concrete implementation of the Mesos scheduling API that enables the deployment of tasks to the edge devices. Using the basic operations supported in the Mesos API, EdgeSphere's scheduler connects to the master and accepts offers from the connected Mesos workers. EdgeSphere provides additional functionality beyond what is available in the Mesos scheduler driver, including deploying tasks based on desired attributes of the edge devices, persisting tasks to enable automatic restart, and pushing executable code to edge devices that are not located within the same data center as the master. Each of these features will be discussed in detail in this section.

In Edgesphere, the deployment of tasks is handled by pushing executable code to the workers on the distributed remote edge devices. Users can deploy tasks through EdgeSphere Dashboard (see Fig. 7(a)) by uploading the code through a web portal, selecting the type of application, parameters for executing code, and choosing the attributes of the slaves. EdgeSphere scheduler will match the attributes and application type to the running workers to determine where to deploy the code. For example, the user can choose to deploy an application to only the devices that have a particular sensor (e.g. accelerometer or gps). As described earlier, the EdgeSphere Agent running on the gateways publishes the capabilities of the attached devices as attributes to Mesos. This allows the EdgeSphere scheduler to select the appropriate edge devices to deploy the tasks. When the edge devices are selected, the code is downloaded from the EdgeSphere server to the edge device from the executor. The application zipped file is then unpacked on the local device and ready to be executed.

Once the tasks are deployed to the edge devices, they run inside a "sandbox" on the worker called an executor. EdgeSphere includes executors for Java, Python, Node.js, JavaScript, Shell scripts, and Groovy application tasks. These executors are implementations of the Executor interface from the Mesos Java API and can be easily extended to add more application types. When tasks stop running, a notification is sent back to the EdgeSphere scheduler so that the appropriate action is taken as decribed in the next section.

### D. Handling Node Failures

In EdgeSphere, the high availability of the edge gateways is critical to allow continuous operation of analytics processes running at the edge. In the event of a gateway failure, EdgeSphere must ensure that the running processes are restarted or reconnected to the framework when the gateway agent stops running or restarts. While the Mesos master provides failover capabilities for the master node running in the cloud, the gateway processes are managed by the EdgeSphere scheduler. For IoT analytics tasks deployed at the edge to run continuously, EdgeSphere has the support of automatic agent and task restart capabilities on the scheduler and gateway. Mesos keeps a list of registered agents and tasks in memory, but does not provide persistent storage capabilities on the master. Therefore, when the framework or master restarts, the list of previous tasks and workers will not be available. EdgeSphere adds a persistent database on the cloud service to keep track of agents and tasks when they are created. In case of failure, EdgeSphere retrieves the task information from the database and automatically restarts the tasks. The procedure below describes the steps taken to ensure the high availablity of IoT tasks and how to perform task recovery:

1. When an agent is disconnected (after a timeout period) or fails, a "Task Lost" message is sent to the EdgeSphere framework.
2. The framework retrieves the lost task from the persistent storage and places the task in the scheduling queue.
3. A cron task on the EdgeSphere gateway periodically checks if the agent is running periodically, and restarts the agent when it is stopped.
4. A cron task on the EdgeSphere gateway detects if the gateway has network connectivity or if its IP address has changed. If the IP address has changed, the agent re-registers with the master. If the network is disconnected, it will attempt to reconnect.
5. Once the agent is restarted, the queued task will be automatically redeployed.

The following steps are taken when the running task stops:
1. When a task fails, a "Task Failure" message is sent to the EdgeSphere framework.
2. The framework notifies the administrator via different means such as by sending an email.
3. The system can be configured to automatically restart failed tasks or require manually restarting from the EdgeSphere dashboard.

As we can see from Fig. 3 and Fig. 4, EdgeSphere shares some concepts with the Aurora framework since both are built on top of Mesos. However, Aurora does not provide advanced features suitable for edge computing and edge applications such as edge devices discovery, flexibility of setting attributes and resources specifications (i.e., abstraction of resources), contextualization, resources management, persistent task recovery, scheduling constraints, proxy agents to deploy applications on constrained edge devices.

### V. IMPLEMENTATION AND PROTOTYPE

#### A. Prototype Description

EdgeSphere has been used for several use cases. Fig. 6 shows a sample deployment of EdgeSphere for worker safety scenario [13, 14] in which the industrial workers carry safety wearables to detect different events such as fall, fatigue, temperature, heartbeat, heat exposure, etc. The edge gateway, deployed in Raspberry Pi 3 [12] includes a device connection agent in charge of connecting to the sensor tag (e.g., TI LaunchPad) and is worn by the workers. The beacons are used to locate the workers based on triangulation approach. Safety information (e.g., level of fatigue, heat exposure rate) are collected from the sensors the worker carries. The collected data are pre-processed at the edge devices (i.e., gateway) and advanced cognitive-based analytics are run against the data. Hence, the entire raw sensor data are not sent to the cloud. For the sake of the validation, processed and aggregated data are also stored to the cloud (IBM Bluemix) by using IBM Watson IoT Platform [10, 11] for further analysis and report generation as well as to compare the accuracy of the insights for analytics deployed at the edge versus when they are deployed in the cloud. IBM Watson IoT Platform is a messaging hub used to isolate the back-end analytic systems, running in the cloud, from handling connections from potentially thousands to millions of edge devices. In the scenario, the workers are notified in real-time about any anomaly or event. If the worker is in critical situation, safety agent can be urgently sent to help or assist the worker.

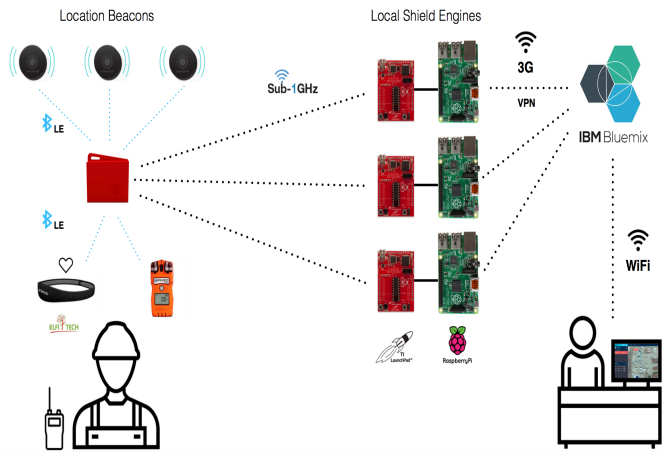

Fig. 6. Prototype architecture for the worker safety use case.

#### B. EdgeSphere Dashboard

Fig. 7 shows the dashboard for the deployment of IoT applications. The application type specifies the runtime for deploying the application; the current types supported are: Java, Python, Node.js, JavaScript, and shell script. The user specifies the main class or script to be executed, the arguments (if needed), and the number of tasks to be executed. The applications are usually a collection of resources (code) provided as files or archives (e.g., tarball or jar). The application requirements in term of resources is also provided for matching with resource offers from the Mesos master. The application resources are uploaded to the selected Mesos agents (for manual selection) or automatically based on the resource offer, the archive is extracted, required package or dependencies are installed, and the application executed. By specifying these attributes as inputs, EdgeSphere provides fine-grained control to satisfy the requirements of the edge applications. Depending on the device type (e.g., specified by its operating system), the user can select attributes from a list of sensors available on the devices, for example accelerometer, gyroscope, magnetometer, and motion.

(a) Dashboard: Deployment View

(b) Dashboard: List of Tasks

Fig. 7. EdgeSphere Dashboard.

## VI. CONCLUSION

In this paper, we have proposed a novel architecture, called EdgeSphere, for deployment and management of advanced analytics and IoT applications running in edge computing environments. EdgeSphere automates the orchestration and deployment of distributed IoT applications while guaranteeing efficient usage of resources, scalability, and fault-tolerance. Resources and attributes of constrained edge devices are abstracted and exposed to a scheduler rather than having access to raw resources on the devices. It is built on top of Apache Mesos and allows extending usage of Mesos on constrained edge devices. Resources abstraction is introduced in EdgeSphere in which the Mesos agent acts as the proxy for the edge devices and is in charge of advertising their resources and capabilities. A discovery mechanism is used for peer edge devices to discover each other and enable support of composite services. EdgeSphere has been built to hide all complexity from the end-user deploying IoT applications. Its performance has been tested along industry grade platform (IBM Watson IoT Platform). A prototype of EdgeSphere has been presented for a worker safety scenario in this paper. However, EdgeSphere has been used in various use cases across industries. Moreover, issues and challenges for deploying and managing with frameworks based on Apache Mesos have been presented.

ACKNOWLEDGMENT

This research was sponsored by the U.S. Army Research Laboratory and the U.K. Ministry of Defence under Agreement Number W911NF-16-3-0001. The views and conclusions contained in this document are those of the authors and should not be interpreted as representing the official policies, either expressed or implied, of the U.S. Army Research Laboratory, the U.S. Government, the U.K. Ministry of Defence or the U.K. Government. The U.S. and U.K. Governments are authorized to reproduce and distribute reprints for Government purposes notwithstanding any copy-right notation hereon.